\documentclass[aps,prl,twocolumn,superscriptaddress,showpacs]{revtex4}
\usepackage{epsfig,graphicx}
\usepackage{indentfirst}
\usepackage{amsfonts,amssymb,latexsym,amsmath,enumerate,amsthm}
\usepackage{bm}
\usepackage{color}
\def\rot{{\rm rot}}

\def\Im{{\rm Im}}

\newcommand{\be}{\begin{equation}}
\newcommand{\ee}{\end{equation}}
\newcommand{\bea}{\begin{eqnarray}}
\newcommand{\eea}{\end{eqnarray}}

\begin{document}

\title{Detecting transition radiation from a magnetic moment}

\author{Igor~P.~Ivanov}
    \email[E-mail: ]{igor.ivanov@ulg.ac.be}
\affiliation{IFPA, Universit\'{e} de Li\`{e}ge, All\'{e}e du
6 Ao\^{u}t 17, b\^{a}timent B5a, 4000 Li\`{e}ge, Belgium}
\affiliation{Sobolev Institute of Mathematics, Koptyug avenue 4, 630090, Novosibirsk, Russia}
\author{Dmitry~V.~Karlovets}
    \email[E-mail: ]{d.karlovets@gmail.com}
\affiliation{Tomsk Polytechnic University,
Lenina 30, 634050 Tomsk, Russia}

\date{\today}

\begin{abstract}
Electromagnetic radiation can be emitted not only by particle charges but also by magnetic moments and higher electric and magnetic multipoles.
However experimental proofs of this fundamental fact are extremely scarce. In particular, the magnetic moment contribution has never been observed
in any form of polarization radiation. Here, we propose to detect it using vortex electrons carrying large orbital angular momentum (OAM) $\ell$.
The relative contribution of the OAM-induced magnetic moment, $\ell \hbar \omega/E_e$, becomes
much larger than the spin-induced contribution $\hbar \omega/E_e$, and it can be observed experimentally.
As a particular example, we consider transition radiation from vortex electrons obliquely incident on an interface between a vacuum and a dispersive medium, 
in which the magnetic moment contribution manifests itself via a left-right angular asymmetry. 
For electrons with $E_e = 300$ keV and $\ell = 100-1000$, we predict an asymmetry of the order of $0.1\%-1\%$, 
which could be measured with existing technology. Thus, vortex electrons emerge as a new tool in the physics of electromagnetic radiation. 
\end{abstract}


\maketitle

\textit{Introduction.} --- Radiation of electromagnetic (EM) waves is an inherent property of charges.
In general, there exist two broad classes of radiation: bremsstrahlung and polarization radiation (PR).
The former is produced by accelerating charges, while the latter can be emitted by a uniformly moving charge
but only in the presence of a medium. 
Depending on the medium or target geometry, one distinguishes
different forms of PR:
Cherenkov radiation, transition radiation, diffraction radiation, Smith-Purcell radiation, etc. (see e.g. \cite{Amusia,PRST,JETP}).

EM radiation can obviously be produced not only by charges but also by neutral particles
carrying higher multipoles: electric or magnetic dipoles, quadrupoles, etc.
For example, transition radiation from these multipoles was studied theoretically
in detail e.g. in \cite{GTs}, while Cherenkov radiation of a
magnetic moment was considered e.g. in \cite{Frank}.
It is therefore remarkable that experimental observations of the
influence of the magnetic moment or of any higher multipole on the EM radiation are very scarce
and are limited to very few cases of spin-induced effects in bremsstrahlung (``spin light'') 
\cite{NIMR1984, Br}. In particular, the contribution of the magnetic moment to any kind of PR
has never been detected, and there are not only technological but also fundamental reasons for that.
Compared with radiation from charge, the relative contribution of the spin-induced magnetic moment to PR
is attenuated by $\hbar \omega/E_e \ll 1$, where
$\hbar \omega$ and $E_e$ are the photon and electron energies, respectively. 
But the quantum effects in radiation are of the same order.
Therefore, this contribution simply cannot be self-consistently calculated 
within the standard quasi-classical treatment of PR, in which one neglects
quantum effects. 

Recently created vortex electrons put a dramatic twist on this problem.
Although solutions of Dirac equation with helical wave fronts were known before
\cite{vortex-early}, it was only in \cite{Bliokh-07} that freely propagating vortex electrons were discussed in detail
and practical methods for their creation were proposed.
Three years later, this proposal was brought to life by several experimental groups \cite{vortex-experimental}.
Vortex electrons carry an intrinsic orbital angular momentum (OAM) $\ell$ with respect to their 
average propagation direction, and values of $\ell \sim 100$ have already been achieved. 
The magnetic moment associated with OAM is correspondingly large, 
$\mu \approx \ell \mu_B$, where $\mu_B = e\hbar/2mc$ is the Bohr magneton. 
One then enters the regime in which the OAM-induced magnetic moment contribution to PR
is only moderately attenuated, 
$\propto\ell \hbar \omega/E_e \lesssim 1$, 
remaining much larger than quantum effects. 
This improves chances to detect this elusive effect and, at the same time,
makes its quasiclassical calculation self-consistent.
This contribution can be predicted, and its observation would be the first clear evidence of PR by a multipole.

In this Letter,
we propose to measure this contribution in transition radiation (TR)
of vortex electrons with $\ell \gg 1$ obliquely incident on an interface between a vacuum 
and a medium with arbitrary (complex) permittivity $\varepsilon (\omega)$. 
We show that the magnetic-moment contribution manifests itself as a left-right asymmetry 
of the emitted radiation with respect to the incidence plane,
and we predict for electrons with $E_e = 300$ keV and $\ell \sim{\cal O}(1000)$ an asymmetry 
of the order of $1\%$.

\textit{TR from ``charge + magnetic dipole'': Qualitative features.} --- Transition radiation occurs when a uniformly moving charge crosses
an interface separating two media with different permittivities \cite{Ginzburg-Frank}.
The accompanying electromagnetic field reorganizes itself when it crosses the interface, and it
is partly ``shaken off'' in the form of electromagnetic radiation, see \cite{GTs} for many details
of the theoretical description of this process. 

\begin{figure}[!htb]
   \centering
\includegraphics[width=6cm]{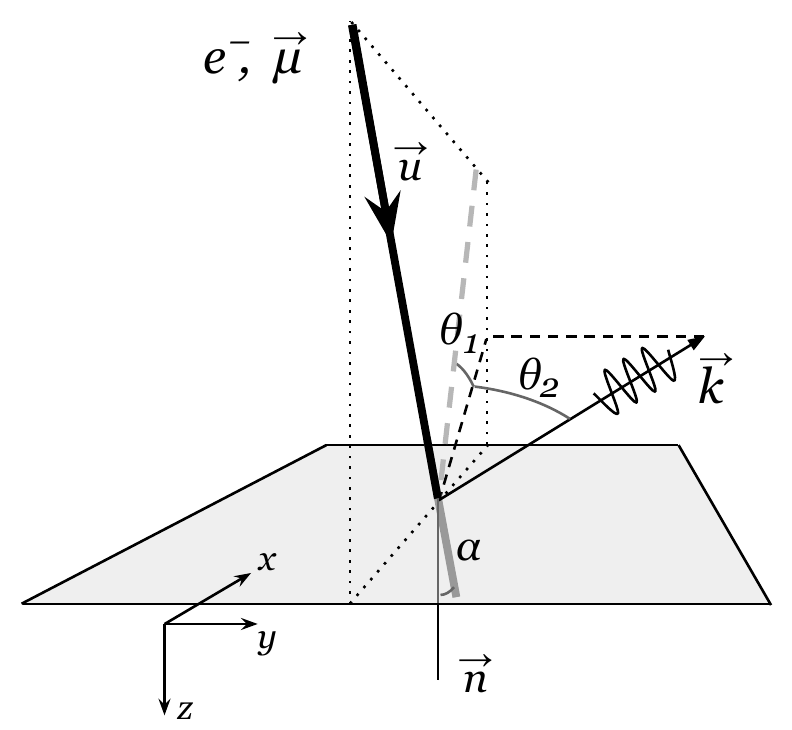}
\caption{Our angle conventions at an oblique incidence with the example of backward TR. The direction of specular reflection is shown by the gray dashed line.}
   \label{fig-geometry}
\end{figure}

Consider first a point-like charge $e$ with no magnetic moment obliquely incident on a flat interface
separating a vacuum from a medium of permittivity $\varepsilon(\omega)$, Fig.~\ref{fig-geometry}.
The angle between the particle trajectory and the normal to the interface 
is $\alpha$. The direction of the emitted photons can be described by two ``flat'' angles:
$\theta_1$ lying in the incidence plane and measured from the direction of specular reflection,
and $\theta_2$ describing an out-of-plane deviation.

TR is mostly emitted into two prominent lobes near the ``forward'' (along the particle velocity) and ``backward'' (i.e. specular) directions,
which are symmetric in $\theta_2$ and have an angular spread of $\sim 1/\gamma = \sqrt{1-\beta^2}$.
The spectrum of TR photons is mostly shaped by the dispersion of the medium, $\varepsilon(\omega)$. 
It stays roughly flat up to $\gamma \omega_p$ \cite{GTs}, with the plasma frequency $\omega_p$ around $10-30$ eV for many materials, and rapidly
decreases above it, thus making the ratio $\hbar \omega/E_e \approx \hbar \omega_p/m_e \sim 10^{-5}$.

TR from a pointlike magnetic moment has also been studied in detail, see e.g. \cite{GTs}.
The main change of the TR from a longitudinally oriented
pointlike magnetic moment $\mu = \ell \mu_B$ with respect to the TR from a charge 
can be anticipated from the comparison of the respective currents:  
${\bm j}_{\mu} = c\ \rot [{\bm \mu} \delta({\bm r} - {\bm u}t)]/\gamma$ vs. ${\bm j}_e = e\, {\bm u}\,\delta({\bm r} - {\bm u} t)$
(here and below $\mu$ denotes the magnetic moment in the particle rest frame; 
in the lab frame it is equal to $\mu/\gamma$). 
Curl leads to an extra factor $i\omega/c$ in the Fourier components of the radiation field,
and the relative strength of the magnetic moment PR always bears the following small factor
\be
x_\ell = \ell {\hbar \omega \over E_e}\,.\label{xell}
\ee
The radiated energy contains this factor squared, making the radiation of pure magnetic moments
many orders of magnitude weaker than that of charges. 
Large $\ell$ partially compensates this suppression, but it still remains prohibitively difficult to detect.

Now, in the case of an electron, we deal with both charge and magnetic moment contributions
to TR. Fields from both sources add up, and the radiated energy can contain three terms
\be
dW = dW_e + dW_{e\mu} + dW_\mu\,,\label{dW3}
\ee
describing the radiation energy of charge $dW_e$, that of magnetic moment $dW_\mu$,
and their interference $dW_{e\mu}$. Since $x_\ell$ is very small,
one can only hope to detect the magnetic moment contribution via $dW_{e\mu}$.

This task turns out to be tricky due to a number of reasons. 
First, ${\bm \mu}$ is a pseudovector, therefore $dW_{e\mu}$
must contain the triple product ${\bm e}_k\cdot [{\bm \mu}\, {\bm n}]$,
where ${\bm e}_k$ is the direction of the emitted photon and ${\bm n}$ is the normal to the interface.
This triple product vanishes for normal incidence, while for oblique incidence it changes sign
under $\theta_2 \to -\theta_2$.
Therefore, the interference can be observed only at oblique incidence and only in a differential distribution,
not in the total energy. It will manifest itself in the form of a left-right asymmetry
\be
A = {W_L - W_R \over W_L + W_R}\,, \quad W_{L,R} = \int d\Omega_{L,R}\, {dW \over d\Omega}\,,
\label{asymmetry}
\ee
where $d\Omega_L$ and $d\Omega_R$ refer to two hemispheres lying to the left and to the right of the incidence plane.
Alternative definitions of this asymmetry using a weight function antisymmetric in $\theta_2$ can also be used.

Next, the curl in ${\bm j}_{\mu}$ produces an extra $i$ factor in the Fourier-components.
As a result, the radiation field ${\bm H}^R$ contains the charge and magnetic moment contributions with a relative phase:
${\bm H}^R = {\bm H}_e + {\bm H}_\mu = a + i x_\ell b$.
These two quantities $a$ and $b$ are complex 
due to the complex $\sqrt{\varepsilon}$, but if they have equal phases,
$dW_{e\mu}$ vanishes.
This happens, in particular, in the cases of a transparent medium ($\Im\, \varepsilon = 0$)
and of an ideal conductor ($\Im\, \varepsilon = \infty$).  
Furthermore, it means that this interference is absent for Cherenkov radiation in a transparent medium.
Observation of a non-zero asymmetry requires, therefore, a real medium
with a sizable (but not asymptotically large)  $\Im\, \varepsilon$, which is the case, for instance, for any real metal.

If all these conditions are satisfied, we can expect, very roughly, 
the asymmetry (\ref{asymmetry}) to be of the order of $A \sim x_\ell$.
For the typical experiments with vortex electrons in microscopes,
this amounts to $A \sim {\cal O}(1\%)$ for optical/UV TR from 
electrons with $\ell \sim {\cal O}(1000)$, and a proportionally weaker
asymmetry for smaller $\ell$.

\textit{TR from vortex electrons: quantitative description.} --- 
A vortex electron state is a freely propagating electron described by a wave function
containing phase singularities with non-zero winding number $\ell$.
Such an electron state is characterized, simultaneously, by an average propagation direction
and by an intrinsic orbital angular momentum (OAM) with a
projection $L = \hbar \ell$ on this direction. 
Following the suggestion \cite{Bliokh-07},
vortex electrons with $E_e = 200-300$ keV and $\ell$ up to 100 
were recently created in experiments by several groups \cite{vortex-experimental}.

The simplest example of a vortex state for a spinless particle is given by the Bessel beam state 
\cite{serbo,K-2012}, described by a coordinate wave function
$\psi(r_\perp, \phi_r, z) \propto e^{ik_z z} e^{i\ell \phi_r} J_{\ell}(k_\perp r_\perp)$.
At large $\ell$, it has a narrow radial distribution 
located around $r_\perp \approx \ell /k_\perp$, confirming the
quasiclassical picture of such an electron as a rotating ring of electronic density.
The spin degree of freedom for a vortex electron was accurately treated in \cite{Bliokh-11,K-2012}.
Both spin and OAM induce magnetic moment \cite{Bliokh-11}, 
see \cite{larmor-precession} for a recent theoretical and experimental investigation of these
contributions, but at large $\ell$ 
the spin contribution and spin-orbital coupling
can be neglected leading to $\mu \approx \ell \mu_B$ (in the electron rest frame).

As explained above, large $\ell$ allows for a self-consistent quasi-classical
treatment of TR from an OAM-induced magnetic moment,
in which the magnetic moment effects of the order of $\ell \hbar \omega/E_e$ are retained
while quantum and spin effects of the order of $\hbar \omega/E_e$ are neglected.
One can then approximate a vortex electron with large $\ell$ by a pointlike particle
with charge $e$ and an {\em intrinsic} magnetic moment $\mu$, and 
calculate TR from both sources,
without the need to discern the microscopic origin of $\mu$. 
The only assumption we make is that, in the absence of magnetic monopoles in nature,
the magnetic moment arises from a closed charge current loop, see discussion on this issue in \cite{GTs}.

To control the validity of this approach, we devised 
another quasi-classical model, in which we treat the vortex electron beam
as a very short bunch of a large number of electrons, $N \gg 1$, 
carrying no intrinsic magnetic moment and uniformly moving along straight rays 
passing through a ring of microscopic size $R \ll \lambda$ at a fixed skew angle.
The calculation is then the standard one of coherent TR from a compact bunch
with the only exception that the total charge of the bunch is just $e$ instead of $N e$.
Using the quasiclassical estimate of
the effective emergent OAM $\ell_{\rm eff} = R p \sin\xi/\hbar$,
where $p$ is the electron momentum and $\xi$ is the skew angle,
we checked that the two models lead to quantitative agreement, see the details in \cite{IK-long}.
Below we focus only on the first model.

These models can be applicable to a realistic experimental set-up with vortex electrons,
if certain coherence conditions are satisfied. 
First, the quasiclassical treatment of the electrons as point-like particles
in the transverse space
is valid only if the vortex electrons are focused in a 
spot of a much smaller size than the emitted light wavelength $\lambda$
(focusing vortex electrons to an {\AA}ngstr\"om size spot was achieved in \cite{verbeeck2}).
The same applicability condition requires also that the longitudinal extent of the 
individual-electron wave function is much shorter than $\lambda$.
This extent can be quantified by the self-correlation length of the electron beam, 
which is related to the monochromaticity of the electron beam and
can be measured experimentally by counting the number of fringes
in a diffraction experiment.
The longitudinal compactness condition implies that the monochromaticity
should not be too good. 

Turning to the calculation of the radiation fields, we use the geometric set-up of Fig.~\ref{fig-geometry} and 
write the electron velocity as ${\bm u} = u (\sin \alpha, 0, \cos \alpha)$.
We start with the currents ${\bm j}_e$ and ${\bm j}_\mu$ of the two sources,
find their Fourier components, calculate the partial Fourier transforms of 
the electric fields they generate, ${\bm E}_e$ and ${\bm E}_\mu$, 
and finally extract the radiation field in the wave zone:
\be
{\bm H}^{R} ({\bm r}, \omega) = \left(\frac{2\pi\omega}{c}\right)^2 \frac{\varepsilon - 1}{4 \pi} 
\,\frac{e^{i\sqrt{\varepsilon}r\omega/c}}{r}\, \left[{\bm e}_{k} \times \bm{\mathcal{J}}\right]\,,\label{HR} 
\ee
where
\be
\bm{\mathcal{J}} = \int dz^{\prime} e^{-i z^{\prime} k_z} 
\left[{\bm E}_e ({\bm k}_{\perp}, z^{\prime}, \omega) + {\bm E}_{\mu} ({\bm k}_{\perp}, z^{\prime}, \omega)\right]\,.\label{pol-current}
\ee
Explicit expressions for the fields and a detailed discussion can be found in \cite{JETP}.
We introduced here the ``on-shell'' wave vector in the medium ${\bm k} = {\bm e}_k \omega/c$, where
\be
{\bm e}_k = \sqrt{\varepsilon} 
\left(\begin{array}{c}
\sin \theta_m \cos \phi \\
\sin \theta_m \sin \phi \\
\cos \theta_m
\end{array}\right)
= 
\left(\begin{array}{c}
\sin \theta \cos \phi \\
\sin \theta \sin \phi \\
\pm \sqrt{\varepsilon_\theta}
\end{array}\right)\,,\label{ek}
\ee
and $\sqrt{\varepsilon_\theta} \equiv \sqrt{\varepsilon - \sin^2\theta}$. 
The two expressions in (\ref{ek}) relate the emission polar angle in the medium $\theta_m$ 
with the emission angle $\theta$ in the vacuum. The latter is connected with the ``flat'' angles $\theta_{1, 2}$ 
by $\cos \theta = \cos \theta_2 \cos (\alpha + \theta_1)$. The integration in (\ref{pol-current}) is carried out from $0$ to $\infty$ for backward TR 
(in this case $e_{k, z} < 0$) and from $-\infty$ to $0$ for forward TR ($e_{k, z} > 0$).

The radiation field can be conveniently written in the coordinates related with the photon production plane $({\bm e}_k, z)$. 
The radiation field (\ref{HR}) is orthogonal to ${\bm e}_k$ and therefore has two components:
one that lies in the production plane, $H^R_{in}$,
and one out of that plane, $H^R_{out}$: 
\begin{widetext}
\begin{eqnarray}
H^R_{out} &=& {\cal N}\Big[ s_\theta (1 - \beta^2 c^2_\alpha - {\bm \beta} \cdot {\bm e}_k) \pm 
\beta^2 s_\alpha c_\alpha c_\phi \sqrt{\varepsilon_\theta} +\ i \mu \frac{\omega}{e\gamma c} s_\alpha s_\phi 
\Big (\beta c_\alpha s^2_\theta \mp \beta s_\alpha s_\theta c_\phi \sqrt{\varepsilon_\theta} \pm \sqrt{\varepsilon_\theta}\Big )\Big]\,,\\
H^R_{in} &=& {\cal N}\sqrt{\varepsilon}\left[ \beta^2 s_\alpha c_\alpha s_\phi 
+ i \mu \frac{\omega}{e\gamma c} \left[\beta s_\theta (1 - s^2_\alpha s^2_\phi) - s_\alpha c_\phi\right]\right] \,,
\eea
\end{widetext}
where we used obvious shorthand notations for sines and cosines and introduced a common kinematical
factor ${\cal N}$, which we omit here. 
Note that at normal incidence, $\alpha = 0$, or for in-plane radiation, at $\phi=0$, the charge contributes 
only to $H^R_{out}$, the magnetic moment contributes only to $H^R_{in}$, 
so that there is no interference. On the other hand, the magnetic moment 
makes TR elliptically polarized, which is another subtle effect to be explored \cite{PKS, PK}. 
The upper and lower signs in these expressions correspond to forward and backward radiation, respectively. 
The spectral-angular distributions of the radiated energy can be found from the reciprocity theorem and reads \cite{JETP}:
\begin{eqnarray}
&& \displaystyle \frac{d^2 W}{d\omega d\Omega} \propto 
\Big |\frac{H^R_{out}\,\cos\theta}{\varepsilon \cos \theta + \sqrt{\varepsilon_\theta}}\Big |^2  
+ \Big |\frac{H^R_{in}\,\cos\theta}{\sqrt{\varepsilon} (\cos \theta + \sqrt{\varepsilon_\theta})}\Big |^2\,.
\label{power-method1}
\end{eqnarray}
Substituting here the explicit expressions for the radiation field and sorting out the charge and magnetic moment contributions, 
one can break the expression for the energy into the three parts introduced in Eq.~(\ref{dW3}).

\begin{figure}[!htb]
   \centering
\includegraphics[width=7cm]{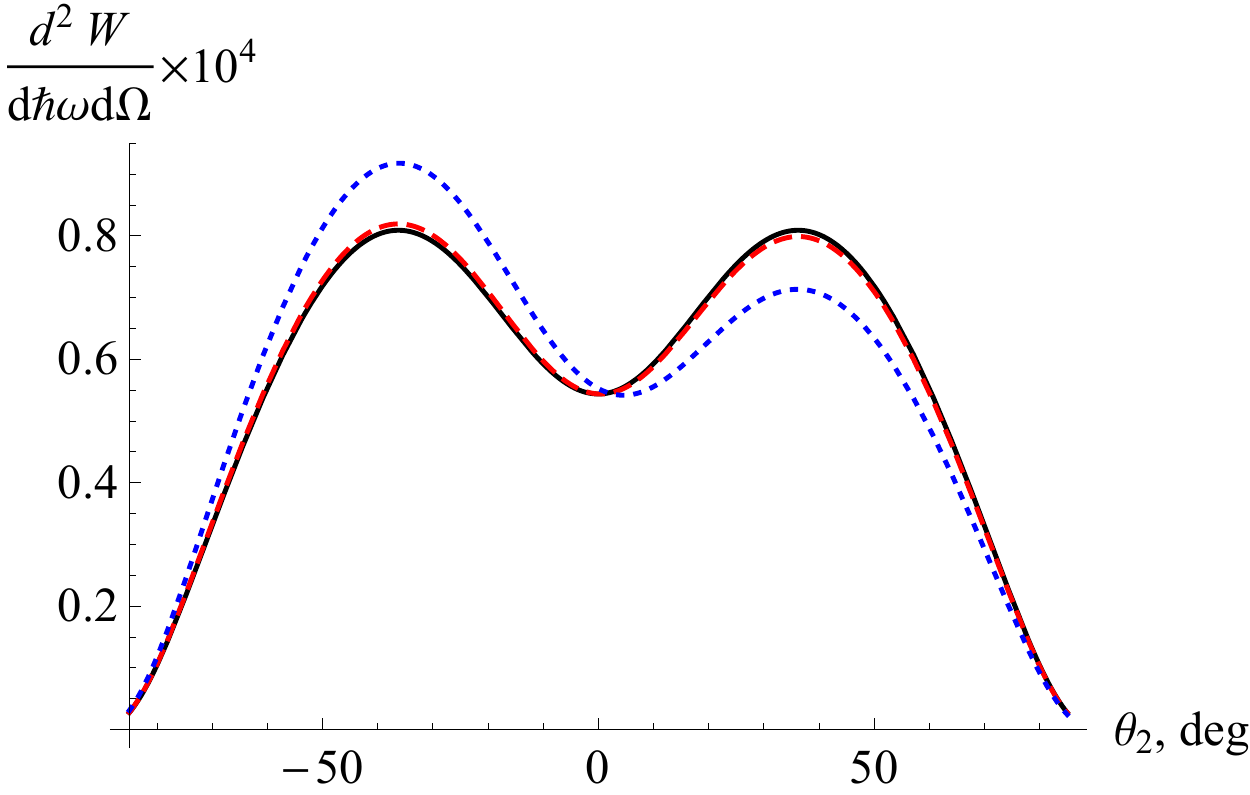}
\caption{(Color online.) Distribution of the forward TR over $\theta_2$ for $\ell = 0$ (black solid line),
$\ell = 1000$ (red dashed line), and $\ell = 10000$ (blue dotted line). Parameters are $\alpha = 70^{\circ}$, $\theta_1 = -40^{\circ}$, $\hbar \omega = 5$ eV.}
   \label{fig2}
\end{figure}

\begin{figure}[!htb]
   \centering
\includegraphics[width=7cm]{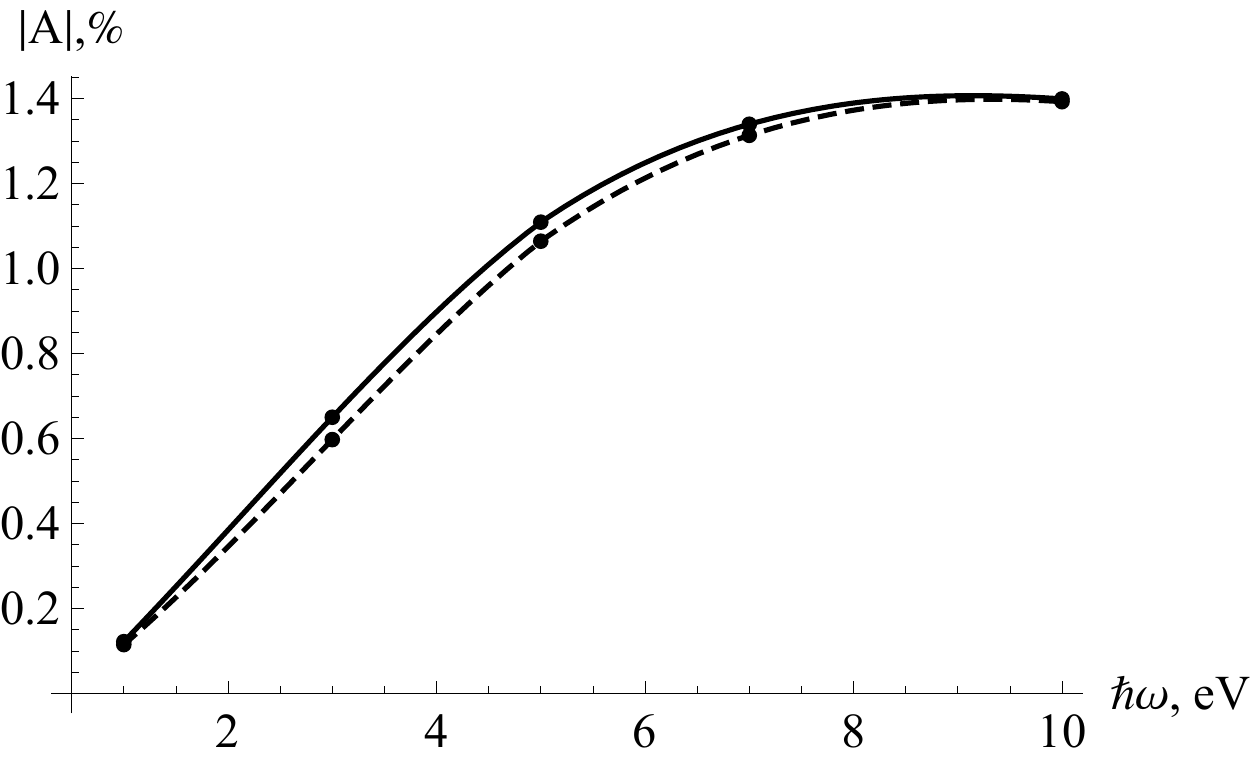}
\caption{The value of the asymmetry $A$ defined in (\ref{asymmetry}) as a function of the emitted
photon energy. The solid and dashed lines correspond to forward and backward TR, respectively. Parameters are $\alpha = 70^{\circ}$, $\theta_1 = -40^{\circ}$, $\ell = 1000$.}
   \label{fig3}
\end{figure}

\textit{Numerical results.} --- In Figs.~\ref{fig2} and \ref{fig3} we show numerical results for 300-keV
electrons incident on an aluminium foil (aluminium permittivity data were taken from \cite{Rakic}).
For non-vortex beams, the angular dependence of TR is $\theta_2$-symmetric, see black curve in Fig.~\ref{fig2}.
Non-zero $\ell$ induces a left-right asymmetry, which becomes huge for $\ell = 10^4$.
For smaller $\ell$, this asymmetry can be extracted via Eq.~(\ref{asymmetry}).
In Fig.~\ref{fig3} we show its magnitude as a function of the photon energy.
The initial rise, $\propto x_\ell \propto \hbar \omega$, slows down above 5 eV due to dispersion, 
which makes the UV-range optimal for detecting the effect, see details in \cite{IK-long}. 
We emphasize that the values of the asymmetry depend rather weakly 
on the target medium (provided it is a metal) and the emission angle $\theta_1$.

\textit{Experimental feasibility.} --- Let us briefly comment on the feasibility of the proposed observation.
The state-of-the-art experiments with vortex electron beams already satisfy 
the coherence requirements. 
The key issue is to obtain large OAM in the first diffraction peak.
So far, $\ell = 25$ has been achieved; a tenfold increase of this value is highly desirable.
Manufacturing such diffraction gratings is challenging but
seems to be within technological limits. 
Alternatively, one can use a novel method for vortex electron generation \cite{verbeeck-monopole}
via the passage of ring-shaped non-vortex electrons through the tip of a
magnetic whisker.
Note also that we do not require the vortex electrons to be in a state of definite $\ell$; 
the effect remains even if OAM is spread over a broad range of values.

Detecting a small asymmetry necessitates large counting statistics.
Our calculations give $n_\gamma \sim {\cal O}(10^{-4})$ TR photons per incident electron,
which can be seen from Fig.~\ref{fig2}.
With a current of 1 nA, easily achievable in vortex electron experiments, 
and a photon detector with 
a quantum efficiency of 10\%, one can expect about $10^5$ photons per second.
With a sufficient integration time, a left-right asymmetry of order $A \sim 0.1\%$ can be reliably detected.

\textit{In summary,} we showed that by studying UV transition radiation from vortex electrons
with large OAM, one can detect for the first time the
magnetic moment contribution to polarization radiation.
For $\ell = 100-1000$ we predict an asymmetry of the order of $0.1\%-1\%$, 
which could be measurable with existing technology. 
Simultaneously, it gives a novel method to measure large OAM
in electron vortex beams. 

\begin{acknowledgments}
The authors are grateful to J.~Verbeeck and members of his team for discussions on the experimental feasibility
of the proposed measurement and to J.-R. Cudell for his remarks on the manuscript.
I.P.I. acknowledges grants RFBR 11-02-00242-a and RF President grant for scientific schools NSc-3802.2012.2.
D.V.K. acknowledges grants of the Russian Ministry for Education and Science within the program ``Nauka'' Nos. 14.B37.21.0911, 14.B37.21.1298,
and the RFBR grant No.12-02-31071-mol\_a.
He also wishes to thank the IFPA group at the University of Li\`ege for hospitality during his visit.
\end{acknowledgments}

\end{document}